\documentclass[twocolumn,showpacs,preprintnumbers,amsmath,amssymb]{revtex4}
\usepackage{graphicx}
\newcommand{\ket}[1]{|#1\rangle}

\begin{document}

\title{A hybrid-exchange density-functional theory study of the electronic structure of $\mathrm{MnV}_2\mathrm{O}_4$: Exotic orbital ordering in the cubic structure}

\author{Wei Wu}\email{wei.wu@ucl.ac.uk}
\affiliation{Department of Electronic and Electrical Engineering and London Centre for Nanotechnology, University College London, Gower Street, London, WC1E 6BT, United Kingdom}

\begin{abstract}
The electronic structures of the cubic and tetragonal $\mathrm{MnV}_2\mathrm{O}_4$ have been studied by using hybrid-exchange density functional theory. The computed electronic structure of the tetragonal phase shows an anti-ferro orbital ordering on V sites and a ferrimagnetic ground state (the spins on V and Mn are anti-aligned). These results are in a good agreement with the previous theoretical result obtained from the local-density approximation+$U$ methods [S. Sarkar, et. al., Phys. Rev. Lett. 102, 216405 (2009)]. Moreover, the electronic structure, especially the projected density of states of the cubic phase has been predicted with a good agreement with the recent soft x-ray spectroscopy experiment. Similar to the tetragonal phase,  the spins on V and Mn in the cubic structure favour a ferrimagnetic configuration. Most interesting is that the computed charge densities of the spin-carrying orbitals on V in the cubic phase show an exotic orbital ordering, i.e., a ferro-orbital ordering along [110] but an anti-ferro-orbital ordering along [$\overline{1}$10].
\end{abstract}

\pacs{ 71.15.Mb, 75.47.Lx, 75.50.Gg, 75.25.Dk}

\maketitle

\section{Introduction}



Many fascinating phenomena in condensed-matter physics were discovered in transition-metal oxides (TMOs) and their closely related compounds such as the first high transition temperature superconductivity in $\mathrm{YBaCuO}_3$ \cite{bednorz1986, wu1987}. An important ordering phenomenon, Orbital ordering (OO) has a long history in the physics of TMOs, back to the 1930s, when the concept of charge ordering in Fe$_3$O$_4$ was proposed \cite{verwey1939}. OO occurs when the orbital degeneracy is lifted essentially by the interaction with the lattice environment \cite{khomskii2003, liechtenstein1995}. The best-known example of OO is the Jahn-Teller (JT) distortion observed in LaMn$\mathrm{O}_3$ \cite{goodenough1955}. Therein the $e_{g}$ degenerate manifold ($d_{z^2}$ and $d_{x^2-y^2}$) in a cubic environment is further splitted by the JT distortion owing to partially filling of electrons. The underlying physics in OO has been well accounted for by the so-called Kugel-Khomskii model \cite{kugel1975}. Most experimental and theoretical works so far have been focused on the system with partially filled $e_g$ orbitals, i.e., $e_g$ active. This type of orbital occupancy can lead to the strongest JT effect because the related $d$-orbitals directly point to ligands and strongly hybridise with the $2p$-orbitals of the anion.

Recently much attention has been paid to vanadium spinels $\mathrm{AB}_2\mathrm{O}_4$ (A = Mg, Co, Fe, Mn, and Cd, etc and B = V) \cite{mun2014, huang2012, kawaguchi2013, adachi2005, sarka2009, nii2013} owing to their fascinating spin and orbital orderings accompanied by complex structural transitions at low temperature. The interplay between structure, orbital, and spin implies that the underlying physics could be very interesting, and that there might be large application potential in these materials. For example, artificially tuning structure by applying external magnetic field can be thought of as a prototype of TMO-based metamaterials or even quantum metamaterials \cite{adachi2005, felbacq2012}. Another unique point for vanadium spinels is that they are $t_{2g}$-active, in sharp contrast to most of the known TMOs showing OO. In this type of compound, the $\mathrm{V}^{3+}$ ions on the B sites form a pyrochlore lattice. The two $d$-electrons on V occupying two $t_{2g}$ orbitals lead to a total spin of $S=1$, resulting in a $t_{2g}$-active OO. The OO in this type of compounds is further complicated by their intrinsic geometrical frustration, which played an important role in OO. In 1939, the charge ordering on the B-site pyrochlore lattice was proposed to cause a sharp increase of the electrical resistivity when cooling below $\sim 120$ K in $\mathrm{Fe}_3\mathrm{O}_4$ (the 'Verwey' transition) \cite{verwey1939}. Anderson was the first to realise the intimate relationship between the proton ordering in the ice, the Verwey charge ordering, and the ordering of Ising spins \cite{anderson1956}. Regardless of the specific material types, geometrical frustration plays a crucial role to determine the spin or charge configurations, by minimising exchange or Coulomb energies. This will in turn trigger many interesting phenomena such as orbital-glass and orbital-ice states \cite{fichtl2005, mulders2009,chern2011}. In addition, the much faster response of electron charges as compared to electron spins will find many potential applications in advanced electronics such as orbitronics \cite{bernevig2005}. 

\begin{figure}[htbp]
\includegraphics[scale=0.5,clip=true]{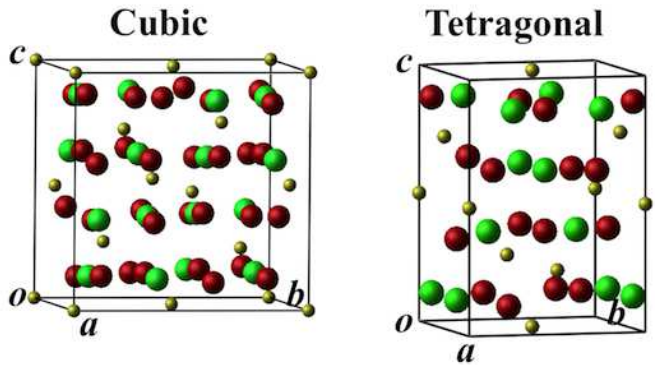}
\caption{ (Color online.) the conventional cells in the cubic (left)  and tetragonal (right) $\mathrm{MnV}_2\mathrm{O}_4$ crystal structures are shown. Mn is depicted as small yellow ball, V as large green ball, and O as large red ball. }\label{fig:mnv2o4}
\end{figure}

$\mathrm{MnV}_2\mathrm{O}_4$ (See Fig.\ref{fig:mnv2o4}), in which the $\mathrm{Mn}^{2+}$ ion is in a $d^5$ half-filled high-spin configuration ($S=\frac{5}{2}$), first experiences a magnetic transition to the collinear ferrimagnetic state at $T= 56$ K, and then a structural distortion at a slightly lower temperature $T = 53$ K, along with a transition to the non-collinear ferrimagnetic state \cite{garlea2008,suzuki2007}. The observed magnetic ordering below $T= 56$ K is ferrimagnetic, i.e., the magnetic moments on $\mathrm{Mn}^{2+}$ and $\mathrm{V}^{3+}$ are anti-aligned. In the structural transition, the symmetry is lowered from cubic to tetragonal. The OO of the ground state in the tetragonal $\mathrm{MnV}_2\mathrm{O}_4$ structure has been shown theoretically \cite{sarka2009} as an $A$-type antiferro-orbital ordering with a propagation vector of (002). 


Previously, the electronic structure and magnetic properties of the tetragonal $\mathrm{MnV}_2\mathrm{O}_4$ have been investigated theoretically using LDA (local-density approximation) + $U$ method \cite{sarka2009}, where an orbital ordering among $\mathrm{V}^{3+}$ was proposed, along with a non-collinear magnetic ordering. Following this first-principles calculation, an analytical modelling \cite{chern2010} has further confirmed that an antiferro-orbital ordering exists in the tetragonal $\mathrm{MnV}_2\mathrm{O}_4$. On the other hand, the electronic structure (especially OO) and magnetic properties of the \textit{cubic} $\mathrm{MnV}_2\mathrm{O}_4$ are yet to be studied in detail. It might be worthwhile to (i) employ a computational method without any adjustable parameters, (ii) take into account the electron correlation properly and (iii) compare the electronic structures of the cubic and tetragonal $\mathrm{MnV}_2\mathrm{O}_4$. Hybrid-exchange functional theory (HDFT), in which the exact exchange is hybridised with generalised gradient approximation (GGA) functional to localise $d$-electrons, has performed well for a wide range of inorganic and organic compounds \cite{chen2014, serri2014}. In this paper, HDFT with PBE0 functional \cite{adamo1998} has been used to study the electronic structure and magnetic properties of the cubic and tetragonal $\mathrm{MnV}_2\mathrm{O}_4$. The results presented here are not only in a good agreement with the previous theoretical and recent experimental studies, but also point to an exotic OO state that mixes ferro-orbital- and anti-ferro-orbital orderings. The rest of the discussion is organised as the following: in \S\ref{sec:computationaldetails} computational details are introduced, in \S\ref{sec:resultsanddiscussions} the calculation results are presented and discussed, and in \S\ref{sec:conclusion} some general conclusions are drawn.  

\section{Computational details}\label{sec:computationaldetails}
The calculations of the electronic structures of the tetragonal and cubic $\mathrm{MnV}_2\mathrm{O}_4$ were carried out by using DFT and hybrid-exchange functional PBE0 as implemented in the CRYSTAL 09 code \cite{crystal09}. The crystal structures of the cubic (space group $Fd\overline{3}m$) and tetragonal (space group $I4_1/amd$) $\mathrm{MnV}_2\mathrm{O}_4$ experimentally determined in Ref.\cite{adachi2005} have been adopted here to perform all the calculations. The basis sets for the atomic orbitals centred on the Mn\cite{peintinger2012}, V\cite{mackrodt1993}, and O\cite{towler1994} atoms, which are designed for solid-state compounds, were used. The Monkhorst-Pack samplings \cite{packmonkhorst} of reciprocal space were carried out choosing a grid of shrinking factor to be $6\times 6 \times 6$ ($6\times 6 \times 5$) in order for a consistency with the ratios among reciprocal lattice parameters for the cubic (tetragonal) $\mathrm{MnV}_2\mathrm{O}_4$. The truncation of the Coulomb and exchange series in direct space was controlled by setting the Gaussian overlap tolerance criteria to $10^{-6}, 10^{-6}, 10^{-6}, 10^{-6}$, and $10^{-12}$ \cite{crystal09}. The self-consistent field (SCF) procedure was converged to a tolerance of $10^{-6}$ a.u. per unit cell (p.u.c). To accelerate convergence of the SCF process, all the calculations have been converged by using a linear mixing of Fock matrices by $30\%$. 

Electronic exchange and correlation are described using the PBE0 hybrid functional \cite{adamo1998}, free of any empirical or adjustable parameter. The advantages of PBE0 include a partial elimination of the self-interaction error and balancing the tendencies to delocalize and localize wave-functions by mixing a quarter of Fock exchange with that from a generalized gradient approximation (GGA) exchange functional \cite{adamo1998}. The broken-symmetry method \cite{noodleman} is used to localize collinear opposite electron spins on atoms in order to describe the anti-ferromagnetic state. 


\section{Results and discussions}\label{sec:resultsanddiscussions}

  



\subsection{Projected densities of states}

\begin{figure*}[htbp]
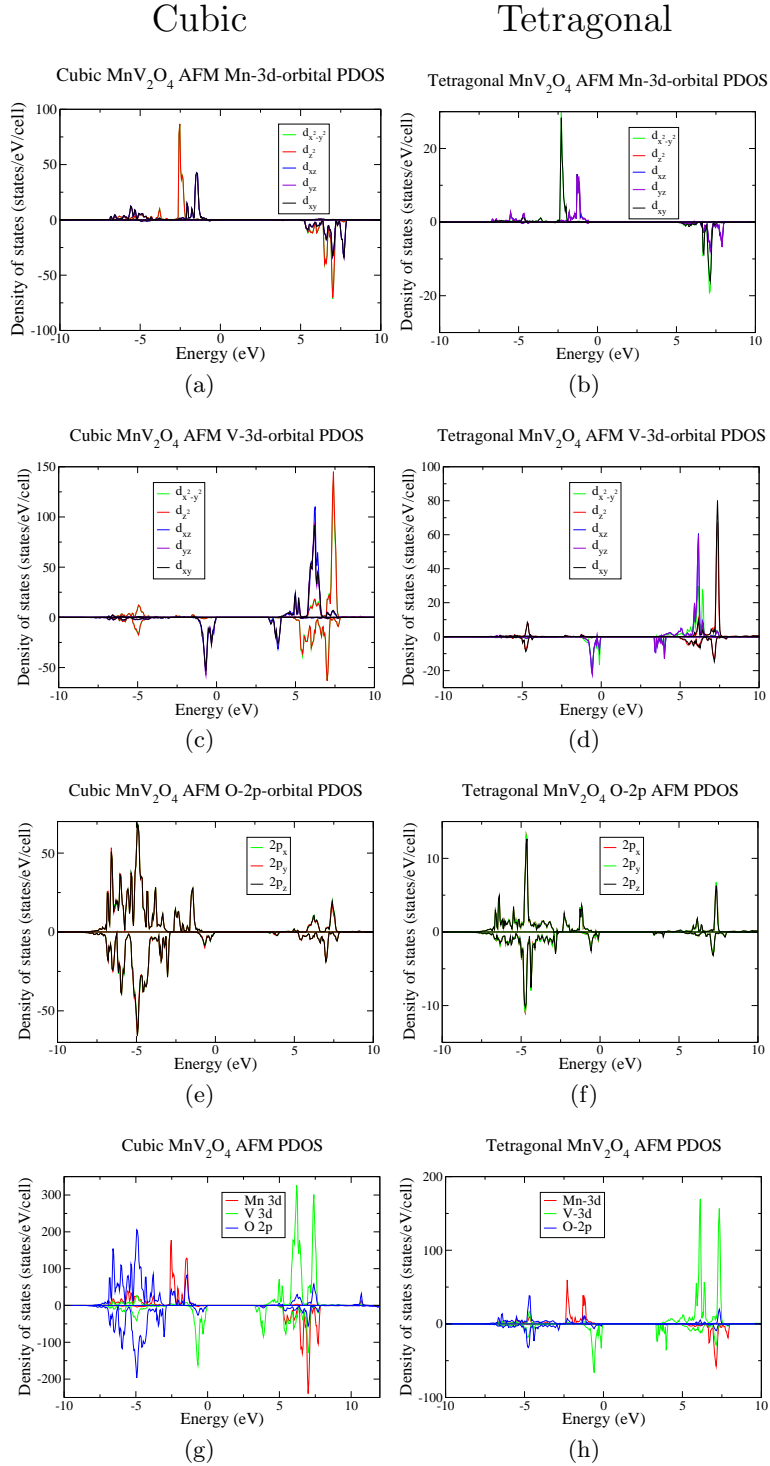

\begin{tabular}{cc}
\centering
 \Large{Cubic}&\Large{Tetragonal} \\
&\\
\includegraphics[scale=0.215]{mnd.eps}&\includegraphics[scale=0.2]{tt_mnd.eps}\\
(a)&(b)\\
&\\
\includegraphics[scale=0.2]{vd.eps}&\includegraphics[scale=0.2]{tt_vd.eps}\\
(c)&(d)\\
&\\
\includegraphics[scale=0.2]{o2p.eps}&\includegraphics[scale=0.2]{tt_2p.eps}\\
(e)&(f)\\
&\\
\includegraphics[scale=0.2]{dptotals.eps}&\includegraphics[scale=0.2]{tt_dptotal.eps}\\
(g)&(h)\\
\end{tabular}
\caption{ (Color online.) The PDOS of the $3d$-orbitals on Mn, the $3d$-orbitals on V, and the $2p$-orbitals on O of the cubic (left) and tetragonal (right) $\mathrm{MnV}_2\mathrm{O}_4$ are shown. The PDOS onto $d_{x^2-y^2}$ is in green, $d_{z^2}$ in red, $d_{xz}$ in blue, $d_{yz}$ in purple, and $d_{xy}$ in black, respectively. For $2p$-orbitals, the PDOS onto $2p_x$ is in green, $2p_y$ in red, and $2p_z$ in black, respectively. In (g) and (h), the total PDOS onto Mn $d$-orbitals is in red, V $d$-orbitals in green, and O $2p$-orbitals in black, respectively.}\label{fig:dos}
\end{figure*}

The projected densities of states (PDOS) of the cubic and tetragonal $\mathrm{MnV}_2\mathrm{O}_4$ structures for the AFM configuration (the spins on Mn and V are anti-ligned) have been shown in Fig.\ref{fig:dos}. The zero energy is aligned with the valence band maximum (VBM). 

In the cubic and tetragonal structures, Mn $3d$ PDOS (Fig.\ref{fig:dos}a and b) shows that the five $d$-orbitals including $d_{x^2-y^2}$, $d_{z^2}$, $d_{xz}$, $d_{yz}$, and $d_{xy}$ are singly occupied, thus giving a spin-$\frac{5}{2}$. From the analysis of Mulliken spin densities and V $3d$ PDOS (Fig.\ref{fig:dos}c and d), the two electrons occupying $t_{2g}$ states can be identified as the origin of the OO both in the cubic and tetragonal $\mathrm{MnV}_2\mathrm{O}_4$. In the cubic phase, the molecular orbitals delocalized among the three $t_{2g}$ states are occupied, whereas in the tetragonal phase, only $d_{xz}$ and $d_{yz}$ are involved. The O $2p$ PDOS (Fig.\ref{fig:dos}e) for the cubic phase are much more delocalized than those for the tetragonal (Fig.\ref{fig:dos}f); this might be due to the elongation of the lattice vector $a$ and $b$ in the tetragonal structure. The O $2p$ PDOS are particularly dominant at $\sim 5$ eV below the VBM for both the cubic and tetragonal structures, which overlap with the rather weak Mn $3d$ and V $3d$ PDOS. The O $2p$ orbitals involved here will give rise to the so-called oxygen bonding states that feature the hybridisation between the $d$-orbitals on transition metals and $p$-orbitals on O atoms. In addition, it can be observed by comparing the spin-up and spin-down PDOS, that the O $2p$ states have been strongly spin-polarized by the magnetic moments on Mn and V. The PDOS onto the V sites have also been compared with the previous results obtained by the LDA$+ U$ \cite{sarka2009}, which suggests that there is a qualitative agreement between them except that the DFT band-gap computed here ($\sim 3$ eV) is larger than the previously computed one ($\sim 1.1$ eV) \cite{sarka2009} and the observed ($\sim 1.1$ eV)\cite{kisma2013}. The computed O $2p$ PDOS are in good agreement with the intensities measured in the recent $K$-edge x-ray absorption and emission spectra; the combination of the theoretical work and experiments will be published in a forthcoming paper \cite{chen2015}.

The computed crystal-field splitting between $t_{2g}$ and $e_g$ states of $\mathrm{Mn}^{2+}$ and $\mathrm{V}^{3+}$ in the cubic $\mathrm{MnV}_2\mathrm{O}_4$ can be read from Fig.\ref{fig:dos}a, which is $\sim 1$ eV. On the other hand, in the tetragonal $\mathrm{MnV}_2\mathrm{O}_4$, for both $\mathrm{Mn}^{2+}$ and $\mathrm{V}^{3+}$, $d_{x^2-y^2}$ and $d_{xy}$ are closely aligned while the other three $3d$-orbitals overlap well, as shown in Fig.\ref{fig:dos}b. This picture is consistent with the previous calculations reported in Ref.\cite{sarka2009}. The crystal-field splitting between these two groups is $\sim 1$ eV. The on-site Coulomb interaction $U$ can be approximated by observing the gap between lower Hubbard $d$-bands and uppers ones ($\sim 5$ eV for Mn site, and $\sim 7$ eV for V), which is much larger than the crystal-field splitting computed here. This will result in a high-spin state for $\mathrm{Mn}^{2+}$ and $\mathrm{V}^{3+}$ ions, which is consistent with the previous experiments \cite{adachi2005}. 

\subsection{Exchange interactions}
The magnetic structure could be much more complicated owing to the exchange interaction between the nearest-neighbouring Mn (V) atoms \cite{nii2013} and the spin-orbit coupling (SOC), which will cause the non-collinear magnetic ordering. However, this topic is beyond the scope of this paper that is focused on the electronic structure and the collinear magnetic ground state. The Mulliken spin densities on the Mn and V sites for the cubic (tetragonal) $\mathrm{MnV}_2\mathrm{O}_4$ are $4.8 \ (4.7) \mu_B$ and $-2.0 \ (-1.9) \mu_B$, respectively. They are close to the expected values, i.e. $5 \mu_B$ for Mn and $-2 \mu_B$ for V (anti-aligned). For the cubic (tetragonal) structure, the computed total energies for a conventional cell with the spins on Mn and V aligned (ferromagnetic) is higher than the anti-aligned (AFM) by $937$ ($754$) meV. This in turn gives an approximate exchange interaction of $\sim 38$ meV (divided by the factor of $4\times S_{\mathrm{Mn}}\times S_{\mathrm{V}}$=20, where $4$ is the number of Mn-V pairs in the unit cell for the tetragonal structure). Similarly we can estimate the exchange interaction for the cubic structure, which is $\sim 23$ meV. On the other hand, the exchange interaction between the Mn and V spins can also be estimated by using the super-exchange formalism, where $t$ can be quantified in the PDOS, which is $\sim 0.5$ eV for Mn and $\sim 1$ eV for V. By using these hopping integrals, the exchange interaction can be estimated as $\frac{t_{\mathrm{Mn}}t_{\mathrm{V}}}{\overline{U}}$ ($\overline{U}=\frac{U_{\mathrm{Mn}}+U_{\mathrm{V}}}{2}$), which is $\sim 59$ meV. This is in the same order as that calculated from the total energy differences aforementioned (here an average value is taken for on-site Coulomb interaction). A detailed discussion about the exchange interaction, taking into account all the $d$-orbitals on both Mn and V would be great, but beyond the scope of this paper.

\subsection{Orbital ordering}
\begin{figure*}[htbp]
\centering
\includegraphics[scale=1.0]{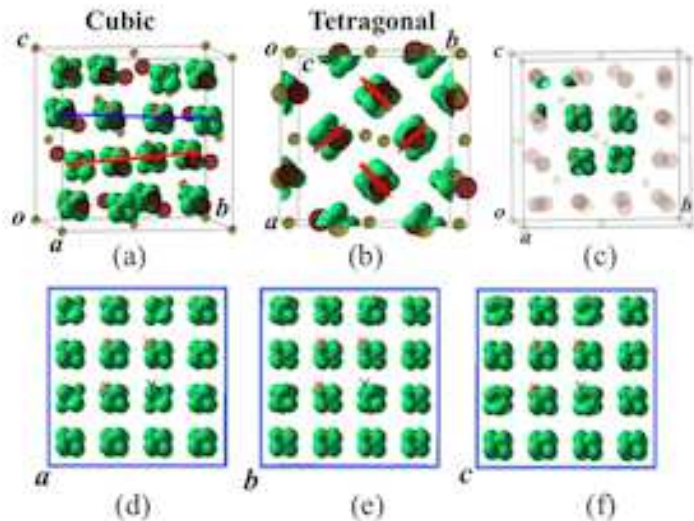}
\caption{ (Colour on line.) The spin densities on V of the AFM state in the conventional unit cell of the cubic (a) and tetragonal (b) structures of $\mathrm{MnV}_2\mathrm{O}_4$ are shown. In the tetragonal an anti-ferro-orbital ordering is found. In contrast, in the cubic structure, the rotation of the orbital orientation illustrates an exotic OO, in which an ferro-orbital ordering is along [110], while anti-ferro-orbital ordering along [$\overline{1}$10]. This exotic OO is further confirmed by the calculation from B3LYP functional; the resulting spin densities are shown in (c). The B3LYP spin-density calculation is performed in a unit cell, but shown in a conventional cell.  The views from [100] ($\vec{a}$), [010] ($\vec{b}$), [001] ($\vec{c}$) lattice directions are shown in (d), (e), and (f), respectively. The isovalue is chosen as $0.04 e/\AA^3$. }\label{fig:spin}
\end{figure*}


The OO in the cubic and tetragonal $\mathrm{MnV}_2\mathrm{O}_4$ can be illustrated by the spin densities on V (see Fig.\ref{fig:spin})-- the difference between the charge densities of spin-up and spin-down orbitals (essentially the charge densities of the spin-carry orbitals). In the tetragonal structure, the relative orientation rotation of neighbouring orbitals is illustrated by the red arrows in Fig.\ref{fig:spin}b; this is consistent with the previous calculations presented in Ref.\cite{sarka2009} in which an anti-ferro-orbital ordering (AFOO) has been predicted. The orbital orientations for the nearest-neighbouring (NN) orbitals are perpendicular to each other (defined as AFOO). In sharp contrast, for the cubic structure the NN orbtibals (labeled by X and Y in Fig.\ref{fig:spin}a) are organized in an exotic way, in which the orbitals on one of the four symmetry-inequivalent V atoms in the unit cell has a different orbital orientation (labeled by Y), perpendicular to the others (labeled by X). The orbital orientations corresponding to Fig.\ref{fig:spin}a are further illustrated in Fig.\ref{fig:spin}d, e, and f, which are the views of the spin densities from the lattice direction [100] ($\vec{a}$), [010] ($\vec{b}$), and [001] ($\vec{c}$), respectively. This leads to an exotic OO: a ferro-OO (FOO) along [110] (blue arrow in Fig.\ref{fig:spin}), whereas an AFOO along [$\overline{1}$10] (red arrow in Fig.\ref{fig:spin}). This OO can probably be attributed to (i) the crystal field environment formed by neighbouring oxygen atoms and (ii) the Coulomb interaction between $d$-orbitals. This interesting OO has been further confirmed by the calculations with a different type of exchange-correlation functional, B3LYP \cite{b3lyp}, which relies on the empirical parameters for mixing exact exchange and GGA exchange functional. The spin densities on V predicted by using B3LYP functional are shown in Fig.\ref{fig:spin}c, in which FOO is along [110] and AFOO [$\overline{1}$10] (all the atoms are made partially transparent in order to illustrate OO). 

\begin{figure}[htbp]
\includegraphics[scale=0.6]{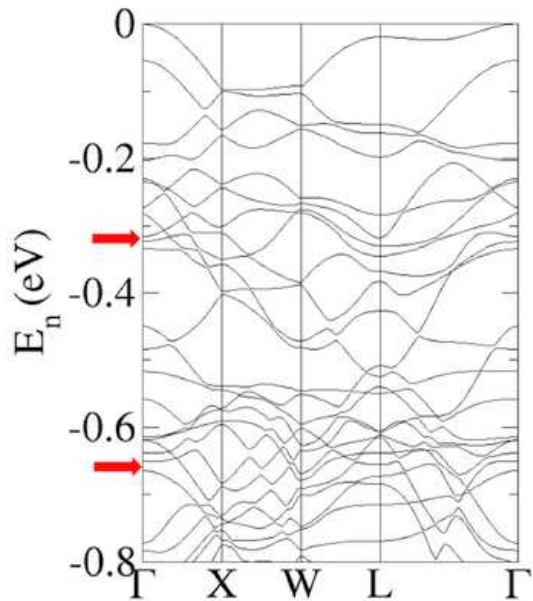}\\
\caption{ (Colour on line.) The spin-down band structure (from -0.8 to 0 eV) of the AFM configuration in the cubic MnV$_2$O$_4$ is shown. Note that the two red arrows refer to VB-8 and VB-9, and VB-19 and VB-20, respectively.}\label{fig:cubicdown}
\end{figure}

To see the orbitals involved in the OO of the cubic MnV$_2$O$_4$ more clearly, the atomic orbital compositions of the bands (see Fig.\ref{fig:cubicdown}) at the $\Gamma$-point that contribute most to the two peaks at $-0.26$ eV and $-0.64$ eV in the V spin-down PDOS (see Fig.\ref{fig:dos}c) have been investigated. The bands VB$-8$ and VB$-9$ contribute the most to the peak at $-0.26$ eV, whereas VB$-19$ and VB$-20$ to the one at $-0.64$ eV. Although the atomic coefficients for $d$-orbitals are slightly different among V atoms, approximately the most dominant linear combinations of  $d$-orbitals for each V atoms contained in these bands read,

\begin{eqnarray}
&&\phi_{\mathrm{VB}-8}=0.2\ket{d_{xz}}+0.1\ket{d_{yz}}+0.2\ket{d_{xy}},\label{eq:vbs1}\\
&&\phi_{\mathrm{VB}-9}=-0.1\ket{d_{xz}}+0.2\ket{d_{yz}}-0.2\ket{d_{xy}},\label{eq:vbs2}\\
&&\phi_{\mathrm{VB}-19}=0.2\ket{d_{xz}}-0.1\ket{d_{yz}}-0.2\ket{d_{xy}},\label{eq:vbs3}\\
&&\phi_{\mathrm{VB}-20}=0.2\ket{d_{xz}}+0.2\ket{d_{yz}}+0.1\ket{d_{xy}}.\label{eq:vbs4}
\end{eqnarray}

\begin{figure*}[htbp]
\begin{tabular}{cccc}
\centering
\includegraphics[scale=0.37]{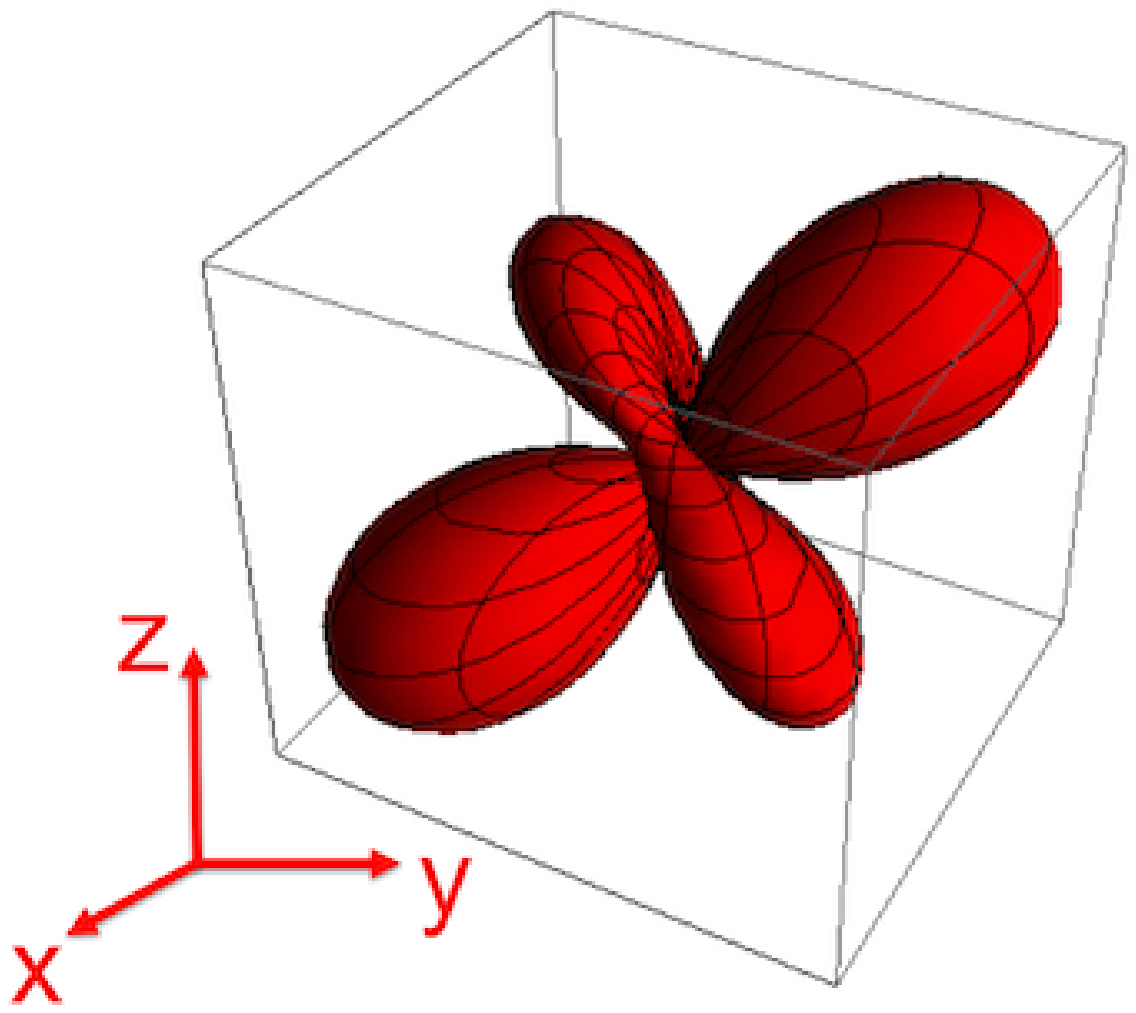}&\includegraphics[scale=0.37]{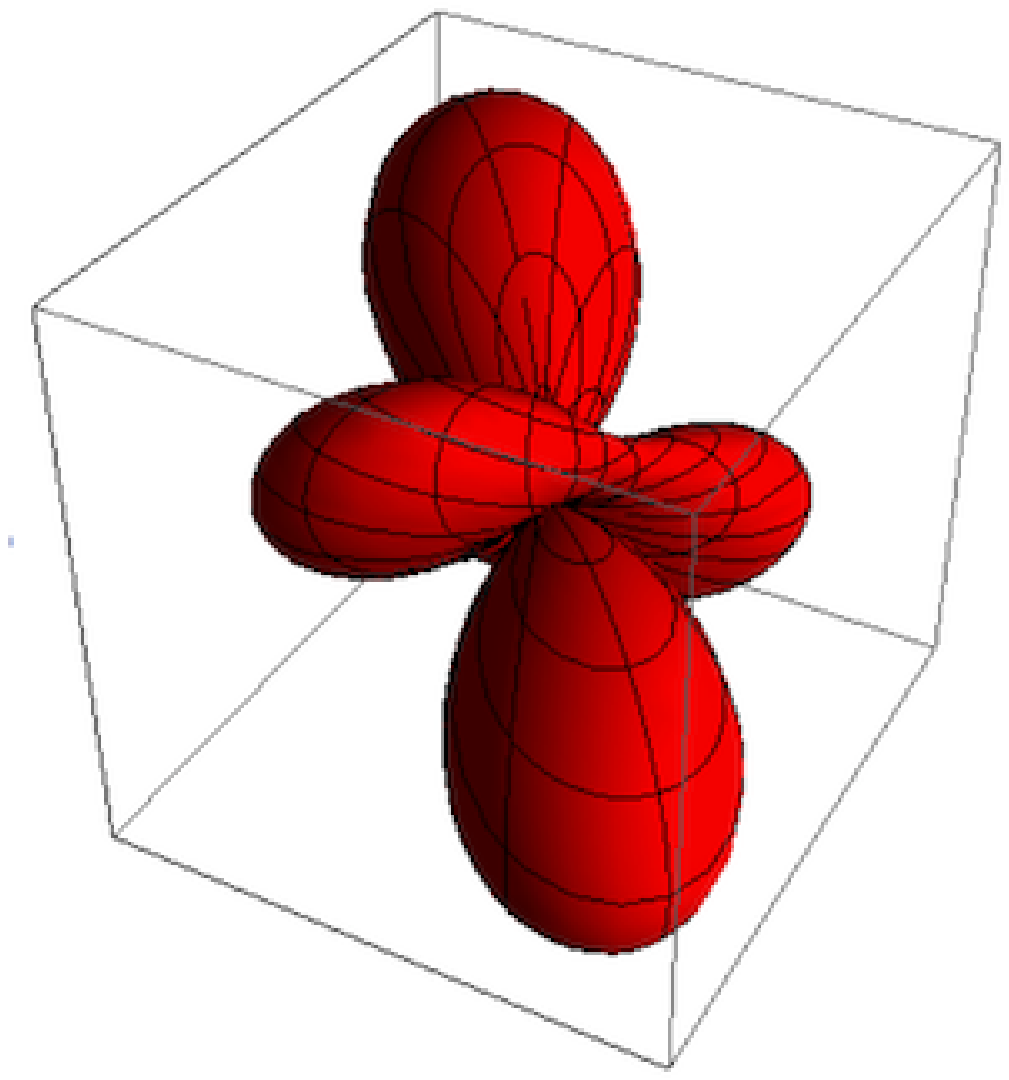}&\includegraphics[scale=0.28]{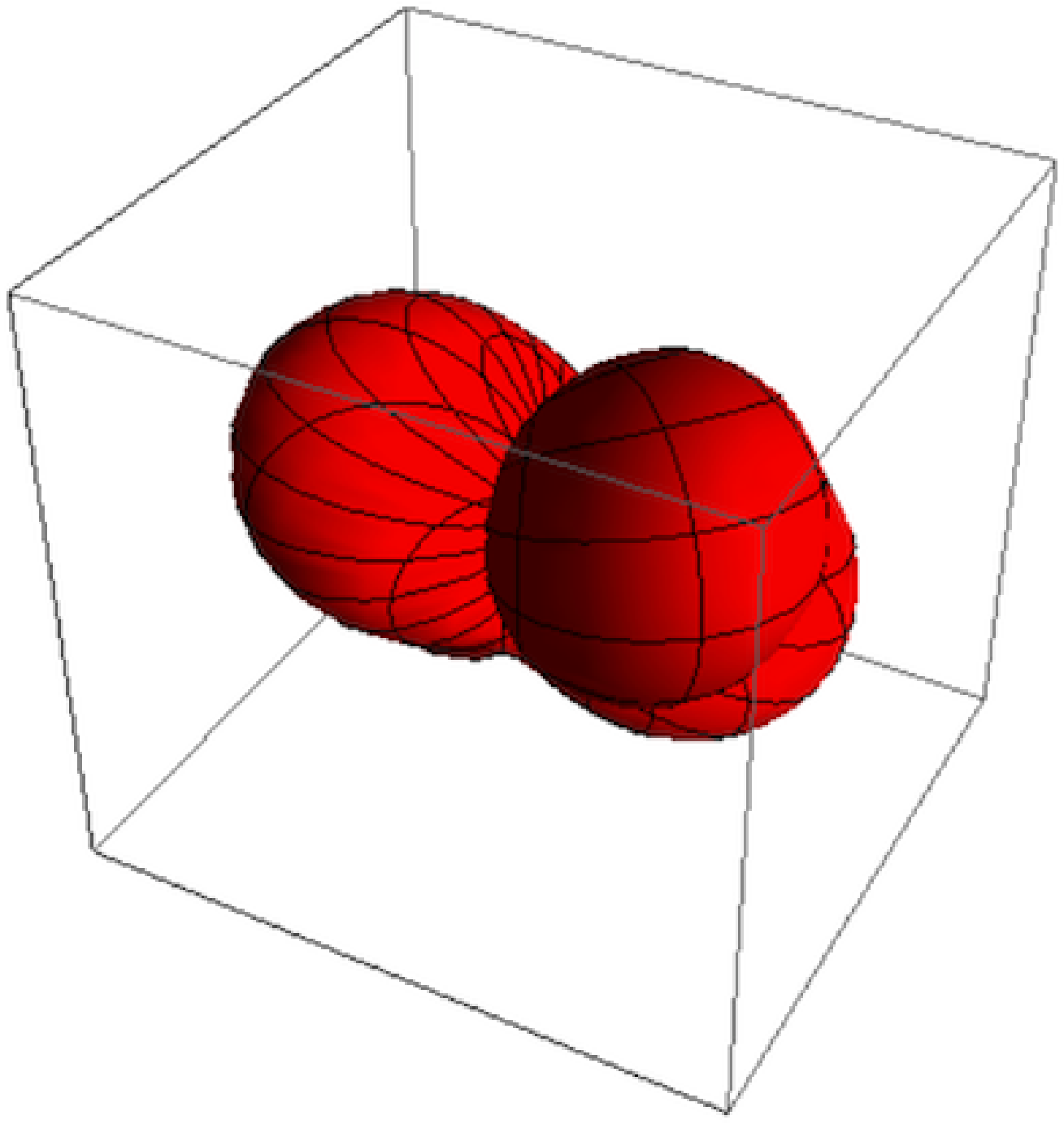}&\includegraphics[scale=0.28]{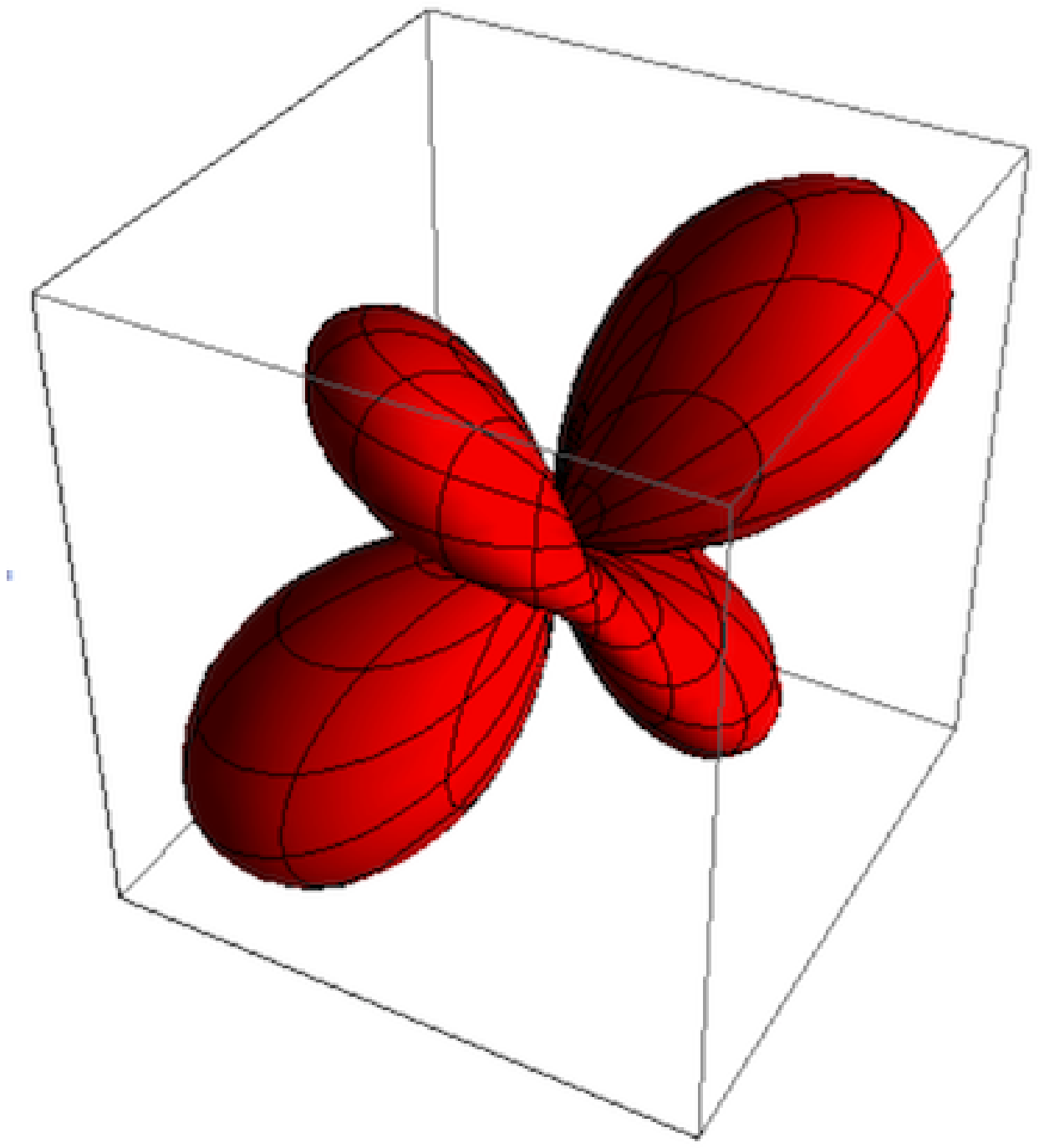}\\
(a)&(b)&(c)&(d)\\
\end{tabular}
\caption{ (Colour online.) The linear combination of harmonic spherical functions are shown. $2Y_{xz}+Y_{yz}+2Y_{xy}$ (corresponding to eq.(\ref{eq:vbs1})) are shown in (a), $-Y_{xz}+2Y_{yz}-2Y_{xy}$ (corresponding to eq.(\ref{eq:vbs2})) in (b), $2Y_{xz}-Y_{yz}-2Y_{xy}$ (corresponding to eq.(\ref{eq:vbs3})) in (c), and $2Y_{xz}+2Y_{yz}+Y_{xy}$ (corresponding to eq.(\ref{eq:vbs4})) in (d). Notice that the linear combination illustrated in (c) changes the orientation of orbitals significantly away from their original ones, which is closely related to the exotic OO shown in Fig.\ref{fig:spin}a.}\label{fig:vbs}
\end{figure*}

The corresponding spherical harmonic functions have been plotted in Fig.\ref{fig:vbs} to see the contributions of $d$-orbitals. From the perspectives of linear combination of atomic orbitals (LCAO), the strong variations of the orbital orientation illustrated in Fig.\ref{fig:spin} and Fig.\ref{fig:vbs}, result from the minimisation of the total energy respective to atomic orbital coefficients. On the other hand, the process of energy-minisation can be seen as the orbital re-orientation driven by Coulomb and crystal-field interactions. 


There are two concepts that are closely related to this exotic orbital ordering, including orbital glass and ice states. The orbital glass state is formed by the randomly orientated orbitals, which result from Jahn-Teller distortion and orbital frustrations in a similar manner to spin frustrations \cite{fichtl2005, mulders2009}. In light that orbital glass states have been observed in the spinel compounds $\mathrm{FeCr}_2\mathrm{S}_4$ and $\mathrm{LuFe}_2\mathrm{O}_4$ \cite{fichtl2005, mulders2009} (which has similar chemical formula to a spinel, but a different structure), this is not completely unexpectable in the cubic $\mathrm{MnV}_2\mathrm{O}_4$. Therefore, these calculations could suggest that at $T= 53$ K, there is not only a structural transition, but also an orbital-ordering transition from the exotic OO illustrated in Fig.\ref{fig:spin} to anti-ferro orbital ordering \cite{sarka2009}, when lowering temperature. In addition, Chern and the collaborators have presented an ice model consisting of triplet orbital variables (such as $p$-orbitals), which is however closely related to orbital-driven many-body phenomena in optical lattice \cite{chern2011}. The analogue of spin-interacting Hamiltonian, an orbital-exchange model describing Coulomb interactions has been used therein. The exotic ordering presented here is closely related to the newly proposed concept of the orbital ice \cite{chern2011}, similar to the spin ice. 

\section{Conclusion} \label{sec:conclusion}
The electronic structures of the cubic and tetragonal $\mathrm{MnV}_2\mathrm{O}_4$ structures have been computed by using hybrid-exchange density functional theory PBE0 and B3LYP. The calculated PDOS of the tetragonal structure has been in a good agreement with previous theoretical results. The most important finding here is that the charge densities of spin-carrying orbitals suggest a possible existence of an exotic orbital ordering: FOO along [110] and AFOO [$\overline{1}$10] in the cubic $\mathrm{MnV}_2\mathrm{O}_4$. The theoretical prediction presented here could be validated in the future experiment. For example,  the different OOs along those two lattice orientations aforementioned could lead to different electron transport properties. More advanced theoretical methods such as dynamical mean-field theory \cite{kotliar2006} is also needed to further understand the electronic structure of $\mathrm{MnV}_2\mathrm{O}_4$. In the ground state the spins on Mn and V are predicted to be anti-aligned, suggesting a ferrimagnetic state for both structures, which is in good agreement with previous theoretical and experimental results. The lower and upper Hubbard bands of Mn and V $d$-electrons have been shown clearly below and above the band gap. The on-site Coulomb interaction can be estimated by using the gap between the lower and upper Hubbard bands, which is $\sim 7$ eV for V and $\sim 5$ eV for Mn, respectively. The transfer integral responsible for the delocalization is in the order of $1$ eV, which should be attributed to a combination of the first-order direct hopping between Mn and V and the second-order effect via O atoms. 
\section{Acknowledgement}
I would like to acknowledge the stimulating discussions with Dr. Leonardo Bernasconi, Dr. Bo Chen, and Dr. Jude Laverock. I also thank Mr. Barry Searle for technical supports and Rutherford Appleton Laboratory for computational resources.


\begin{thebibliography}{99}

\bibitem{bednorz1986} J. G. Bednorz and K. A. MŸller, Z. Phys. B \textbf{64}, 189 (1986).

\bibitem{wu1987} M. K. Wu, J. R. Ashburn, C. J. Torng, P. H. Hor, R. L. Meng, L. Gao, Z. J. Huang, Y. Q. Wang, and C. W. Chu, Phys. Rev. Lett. \textbf{58}, 908 (1987).

\bibitem{verwey1939} E. J. W. Verwey, Nature \textbf{144}, 327 (1939).



\bibitem{khomskii2003} I. D. Khomskii and M. V. Mostovoy, J. Phys. A: Math. Gen. \textbf{36}, 9197 (2003).

\bibitem{liechtenstein1995} A. I. Liechtenstein, V. I. Anisimov, and J. Zaanen, Phys. Rev. B \textbf{52}, R5467(R) (1995).

\bibitem{goodenough1955} J. B. Goodenough, Phys. Rev. \textbf{100}, 564 (1955).

\bibitem{kugel1975} K. I. Kugel and and I. D. Khomskii, Sov. Phys. --- Solid State \textbf{17}, 285 (1975) 

\bibitem{mun2014} E. D. Mun, Gia-Wei Chern, V. Pardo, F. Rivadulla, R. Sinclair, H. D. Zhou, V. S. Zapf, and C. D. Batista, Phys. Rev. Lett. \textbf{112}, 017207 (2014).

\bibitem{huang2012} Y. Huang, Z. Yang and Y. Zhang, J. Phys.: Condens. Matter \textbf{24}, 056003 (2012).

\bibitem{kawaguchi2013} S. Kawaguchi, H. Ishibashi, S. Nishihara, M. Miyagawa, K. Inoue, S. Mori and Y. Kubota, J. Phys.: Condens. Matter \textbf{25}, 416005 (2013).

\bibitem{adachi2005} K. Adachi, T. Suzuki, K. Kato, K. Osaka, M. Takata, and T. Katsufuji, Phys. Rev. Lett. \textbf{95}, 197202 (2005).

\bibitem{sarka2009} S. Sarkar, T. Maitra, R. Valent\'i, and T. Saha-Dasgupta, Phys. Rev. Lett. \textbf{102}, 216405 (2009).

\bibitem{nii2013} Y. Nii, N. Abe, and T. H. Arima, Phys. Rev. B \textbf{87}, 085111 (2013).

\bibitem{felbacq2012} D. Felbacq and M. Antezza ÒQuantum metamaterials: A brave new worldÓ. SPIE Newsroom. (2012). DOI:10.1117/2.1201206.004296

\bibitem{anderson1956} P. W. Anderson, Phys. Rev. B \textbf{102}, 1008 (1956).

\bibitem{fichtl2005} R. Fichtl, V. Tsurkan, P. Lunkenheimer, J. Hemberger, V. Fritsch, H.-A. Krug von Nidda, E.-W. Scheidt, and A. Loidl, Phys. Rev. Lett. \textbf{94}, 027601 (2005).

\bibitem{mulders2009} A. M. Mulders, S. M. Lawrence, U. Staub, M. Garcia-Fernandez, V. Scagnoli, C. Mazzoli, E. Pomjakushina, K. Conder, and Y. Wang, Phys. Rev. Lett. \textbf{103}, 077602 (2009).

\bibitem{chern2011} Gia-Wei Chern and Congjun Wu, Phys. Rev. E \textbf{84}, 061127 (2011).

\bibitem{bernevig2005} B. A. Bernevig, T. L. Hughes, and S-C. Zhang, Phys. Rev. Lett. 95, 066601 (2005). 

\bibitem{garlea2008} V. O. Garlea, R. Jin, D. Mandrus, B. Roessli, Q. Huang, M. Miller, A. J. Schultz, and S. E. Nagler, Phys. Rev. Lett. \textbf{100}, 066404 (2008).

\bibitem{suzuki2007} T. Suzuki, M. Katsumura, K. Taniguchi, T. Arima, and T. Katsufuji, Phys. Rev. Lett. \textbf{98}, 127203 (2007).

\bibitem{chern2010} Gia-Wei Chern, N. Perkins, and Z. Hao, Phys. Rev. B \textbf{81}, 125127, (2010).

\bibitem{chen2014} B. Chen, J. Laverock, D. Newby, Jr., Ting-Yi Su, K. E. Smith, Wei Wu, L. H. Doerrer, N. F. Quackenbush, S. Sallis, L. F. J. Piper, D. A. Fischer, and J. C. Woicik, J. Phys. Chem. C \textbf{118}, 1081 (2014).

\bibitem{serri2014} M. Serri, Wei Wu, L. Fleet, N. M. Harrison, C. W. Kay, A. J. Fisher, C. Hirjibehedin, G. Aeppli, S. Heutz, Nat. Commun. \textbf{5}, 3079 (2014).

\bibitem{adamo1998} C. Adamo and V. Barone, J. Chem. Phys., \textbf{110}, 6158 (1998).

\bibitem{yamada1999} Hiroyuki Yamada and Yutaka Ueda,  J. Phys. Soc. Jpn. \textbf{68}, 2735 (1999).

\bibitem{yamauchi2002} T. Yamauchi, Y. Ueda, and N. M\^{o}ri, Phys. Rev. Lett. \textbf{89}, 057002 (2002).

\bibitem{ma2006} C. Ma, R.J. Xiao, H.X. Yang, Z.A. Li, H.R. Zhang, C.Y. Liang, J.Q. Li, Solid State Communications \textbf{138}, 563 (2006)










\bibitem{crystal09} R. Dovesi, V. R. Saunders, C. Roetti, R. Orlando, C. M. Zicovich-Wilson, F. Pascale, B. Civalleri, K. Doll, N. M. Harrison, I. J. Bush, P. D'Arco, and M. Llunell, CRYSTAL09 User's Manual (University of Torino, Torino, 2009). 

\bibitem{peintinger2012} M. F. Peintinger, D. Vilela Oliveira, and T. Bredow, J. Comput. Chem. \textbf{34}, 451 (2012).

\bibitem{mackrodt1993} W.C. Mackrodt, N.M Harrison, V.R. Saunders, N.L.Allan, M.D. Towler, E. Apra' and R. Dovesi, Philos. Magaz. A \textbf{68}, 653 (1993).

\bibitem{towler1994} M.D. Towler, N.L. Allan, N.M. Harrison, V.R. Saunders, W.C. Mackrodt and E. Apra', Phys. Rev. B \textbf{50}, 5041 (1994).





\bibitem{packmonkhorst} H. J. Monkhorst and J. D. Pack, Phys. Rev. B \textbf{13}, 5188 (1976).


\bibitem{noodleman} L. Noodleman, J. Chem. Phys. \textbf{74}, 5737 (1980).

\bibitem{b3lyp} A. D. Becke, J. Chem. Phys. \textbf{98}, 5648 (1993).








\bibitem{kisma2013} A. Kismarahardja, J. S. Brooks, H. D Zhou, E. S. Choi, K. Matsubayashi, and Y. Uwatoko, Phys. Rev. B \textbf{87}, 054432 (2013).

\bibitem{chen2015} B. Chen, J. Laverock, W. Wu, J. Kuyyalil, D. Newby Jr., K. E. Smith, R. M. Qiao, W. Yang, L. D. Tung, R. P. Singh, and G. Balakrishnan, in preparation.



\bibitem{kotliar2006} G. Kotliar, S. Y. Savrasov, K. Haule, V. S. Oudovenko, O. Parcollet, and C. A. Marianetti, Rev. Mod. Phys. \textbf{78}, 865 (2006).



\end{thebibliography}
\end{document}